\documentclass[journal=jpccck,manuscript=article,layout=twocolumn]{achemso}

\usepackage[utf8]{inputenc}
\usepackage[T1]{fontenc} 
\usepackage{amsmath}
\usepackage{amssymb}
\usepackage{lineno}

\author{Klaus H. Eckstein}
\affiliation[uniwue]{Institute of Physical and Theoretical Chemistry, Julius-Maximilian University W\"urzburg, Germany}
\author{Florian Oberndorfer}
\affiliation[uniwue]{Institute of Physical and Theoretical Chemistry, Julius-Maximilian University W\"urzburg, Germany}
\author{Melanie M. Achsnich}
\affiliation[uniwue]{Institute of Physical and Theoretical Chemistry, Julius-Maximilian University W\"urzburg, Germany}
\author{Friedrich Sch\"oppler}
\affiliation[uniwue]{Institute of Physical and Theoretical Chemistry, Julius-Maximilian University W\"urzburg, Germany}
\author{Tobias Hertel}
\affiliation[uniwue]{Institute of Physical and Theoretical Chemistry, Julius-Maximilian University W\"urzburg, Germany}
\alsoaffiliation{R\"ontgen Research Center for Complex Material Systems, Julius-Maximilian University W\"urzburg, Germany}
\phone{+49 931 3186300}
\email{tobias.hertel@uni-wuerzburg.de}

\title {Quantifying Doping Levels in Carbon Nanotubes by Optical Spectroscopy}


\begin{document}

\begin{abstract}
Controlling doping is essential for a successful integration of semiconductor materials into device technologies. However, the assessment of doping levels and the distribution of charge carriers in carbon nanotubes or other nanoscale semiconductor materials is often either limited to a qualitative attribution of being 'high' or 'low' or it is entirely absent. Here, we describe efforts toward a quantitative characterization of doping in redox- or electrochemically doped semiconducting carbon nanotubes (s-SWNTs) using VIS-NIR absorption spectroscopy. We discuss how carrier densities up to about 0.5 $\rm nm^{-1}$ can be quantified with a sensitivity of roughly one charge per $10^4$ carbon atoms assuming in-homogeneous or homogeneous carrier distributions. 
\end{abstract}

\pagebreak

\section{Introduction}
After nearly two decades of research and in spite of its technological relevance for device applications, the quantitative assessment of carrier concentrations and distributions in redox- or electrochemically doped semiconducting single-wall carbon nanotubes (s-SWNTs) continues to represent a formidable challenge. 

Raman-spectroscopic investigations can reveal doping of SWNTs or graphene from changes in the shape and position of the G-band with a sensitivity on the order of 1 charge per 1000 C-atoms \cite{Rao1997,Ferrari2007,Das2008}. In the case of small diameter s-SWNTs this roughly corresponds to one charge per ten nanometers of tube length which is associated with degenerate doping levels with the Fermi level having been pushed to the nanotube's valence or conduction bands \cite{hertel2018}. However, semiconductor devices such as diodes or FETs generally operate in a considerably less strongly doped regime requiring more sensitive spectroscopic probes for characterization of doping levels. 

Fortunately, the optical spectra of s-SWNTs feature strong exciton bands \cite{Ando1997, Perebeinos2004, Pedersen2004, Spataru2004, Wang2005, Maultzsch2005} as well as clearly separated trion bands in doped samples \cite{Ronnow2009, Matsunaga2011, Santos2011} which both lend themselves as proxies for the assessment of doping \cite{Eckstein2017}.

A trion may be described as a three-particle bound state of an exciton and an extra electron or hole at low doping concentrations  \cite{Ronnow2009, Watanabe2012, Efimkin2017} or as an exciton dressed by the interaction with excitations of the Fermi sea at intermediate doping levels \cite{Efimkin2017}. 

The work presented here discusses the quantitative determination of carrier densities in s-SWNTs from changes in exciton bands in the NIR spectral range. Our findings indicate that both, redox and electrochemical doping schemes are surprisingly similar in how they affect optical spectra. Two models, one based on band-filling and one on exciton confinement for determining carrier concentrations in doped s-SWNTs are compared and are found to yield qualitatively similar results. The findings are also compared to predicted changes of oscillator strengths based on a phase space filling model previously established for measuring exciton Bohr-radii from nonlinear spectroscopy \cite{Schmittrink1985, Luer2009, Mann2016}.

\section{Experimental Section}

Redox doped nanotube samples were prepared from organic (6,5) s-SWNT enriched suspensions of PFO-BPy polymer-stabilized (American Dye Source) CoMoCAT nanotubes (SWeNT SG 65, Southwest Nano Technologies Inc.) as described previously \cite{Hartleb2015,Eckstein2017}. Electrochemical doping was carried out using thin transparent films prepared by vacuum filtration from the same type of colloidal suspensions. Films were contacted by a Pt wire mesh working electrode. All electrochemical experiments were conducted in a 0.1 molar solution of tetrabutylammonium hexafluorophosphate (TBAHFP, Sigma-Aldrich) in dry and degassed tetrahydrofuran under an inert argon atmosphere in a home-built spectroscopy chamber. More details about the electrochemical setup can be found elsewhere \cite{Hartleb2015,Eckstein2017}.

Suspensions with redox-chemically doped s-SWNTs were prepared using the same starting material as for electrochemical measurements. To provide sufficiently good solubility for gold(III)-chloride (Sigma-Aldrich) we used a 5:1 toluene to acetonitrile solvent mixtures for s-SWNTs and redox agent alike.

For electrochemical measurements potentials are measured versus a Pt quasi-reference electrode \cite{Hartleb2015}.

\section{Results and Discussion}

Conventional bulk semiconductors are most commonly doped using substitutional impurities \cite{Sze2007}. In contrast, nanoscale semiconductors such as single-wall carbon nanotubes (s-SWNTs) \cite{Park2012, Yoshida2016, Ferguson2018} or atomically thin two-dimensional semiconductors \cite{Chernikov2015, Peimyoo2014} as well as organic semiconductors are doped using redox chemistry or field-doping \cite{Lussem2013}. 
\begin{figure}[htbp]
	\centering
		\includegraphics[width=8.5 cm]{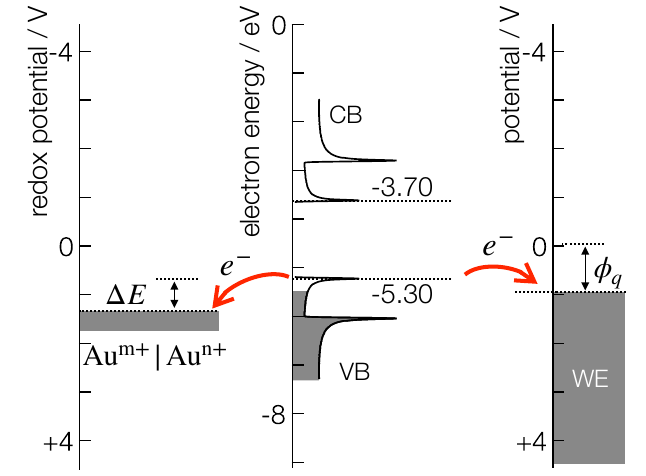}
		\caption{{\bf Schematic illustration of redox- and electrochemical doping.} Redox potential, electron energy and working electrode potential are plotted on the same vertical scale. Redox doping using gold(III)-chloride is shown on the left while field doping in an electrochemical cell is illustrated on the right. The density of states of (6,5) s-SWNTs is shown in the center.}
		\label{fig1}
\end{figure}

The direction of charge transfer during the interaction of carbon nanotubes with gold(III)-chloride solutions is incentivised by the large positive standard redox potentials for the reduction of various Au cations between +1.40 and +1.8\,V \cite{CRC2018} (all standard potentials versus SHE). The redox potential for the reduction of a positively charged (6,5) s-SWNT is about +0.86\,V \cite{Hartleb2015}. The potential difference is thus sufficiently positive such that electrons are readily provided by the $\rm SWNT^+|SWNT$ redox couple, to facilitate the spontaneous reduction of Au cations, even if $\rm AuCl_3$ concentrations are in the micromolar range. This is illustrated schematically in Figure \ref{fig1} where we compare the energetic alignment of (6,5) s-SWNT valence and conduction bands \cite{Hartleb2015} with standard redox potentials for the reduction of different Au cations \cite{CRC2018}. 
\begin{figure}[htbp]
	\centering
		\includegraphics[width=8.5 cm]{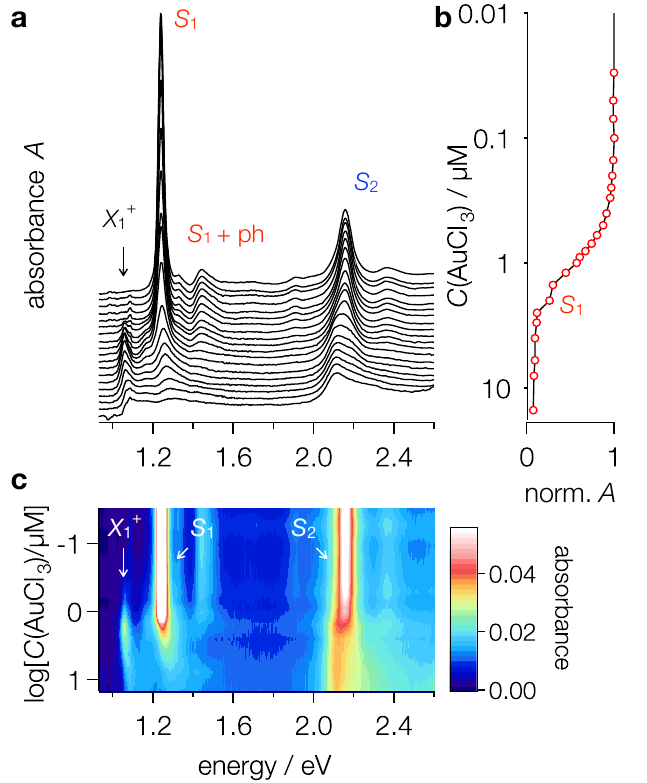}
		\caption{{\bf Absorption spectra of redox-doped s-SWNTs.} {\bf a)} Waterfall plot of absorption spectra from a suspension with (6,5) s-SWNTs redox doped using gold(III)-chloride (intrinsic at the top, heavily doped at the bottom). {\bf b)} Semi-log plot of the first exciton peak absorbance at 1.247\,eV {\it vs.} $\rm AuCl_3$ concentration. c) False color plot of the same data.}
		\label{fig2}
\end{figure}

A series of absorption spectra for a (6,5) s-SWNT suspension in the presence of different concentrations of gold(III)-chloride is shown in Fig. \ref{fig2}a). The peak absorbance of the dominant first subband exciton ($S_1$) at 1.247\,eV is seen to decrease dramatically when the redox concentration is increased as shown on the semi-log plot in Fig. \ref{fig2}b). The same concentration or doping-level dependence is also illustrated in the semi-log false-color plot in Fig. \ref{fig2}c). This clearly reveals the simultaneous intensity decrease of the first and second subband excitons, $S_1$ and $S_2$, respectively alongside the appearance of a positive trion peak at 1.061\,eV designated $X_1^+$. At higher doping levels these quasiparticle excitations disappear and give rise to a broad and featureless absorption signal (H-band) which extends from 1.05\,eV to nearly 2.0\,eV \cite{Hartleb2015,Eckstein2017,Berger2018,Avery2016}.
\begin{figure}[htbp]
	\centering
		\includegraphics[width=8.5 cm]{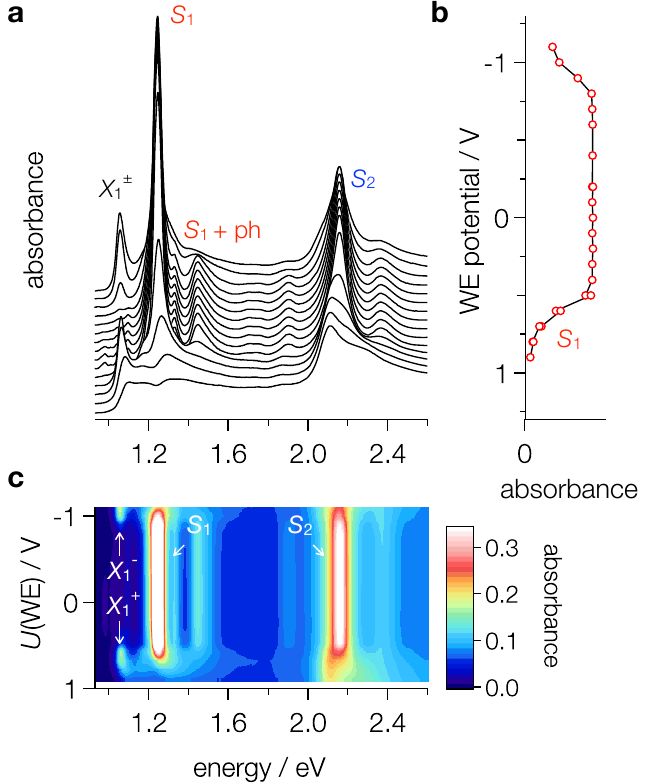}
		\caption{{\bf Absorption spectra of electrochemically doped s-SWNTs.} {\bf a)} Waterfall plot of spectra from a thin film with (6,5) s-SWNTs electrochemically {\it n}-doped (top) and {\it p}-doped (bottom) at potentials between -1.1 and +0.9 V with respect to a Pt pseudo reference electrode. {\bf b)} Peak absorbance of the first subband exciton versus applied potential. {\bf c)} False color plot of the same data.}
		\label{fig3}
\end{figure}

Alternatively, s-SWNTs in contact with a working electrode (WE) can also be doped by applying electrical fields between s-SWNTs and the WE. Field-induced charge transfer is incentivized by $+zF\phi$, the contribution of the electrical potential to the electrochemical potential $\tilde{\mu}$. Here, $z$ is the number of transferred charges, $F$ is the Faraday constant and $\phi$ is the drop of the electrostatic potential at the electrode nanotube interface as induced by the applied circuit voltage $U_{\rm WE}$.

In Fig. \ref{fig3} we have reproduced a series of spectra previously reported by Hartleb {\it et al.} for a thin film of electrochemically field-doped (6,5) s-SWNTs \cite{Hartleb2015}. Fig. \ref{fig3}b) shows the peak absorbance of the first subband exciton at 1.247\,eV as a function of the applied working electrode potential $U_{\rm WE}$. Positive WE potentials lead to {\it p}-doped, negative potentials to {\it n}-doped samples. As for redox-chemically doped s-SWNTs the data reveals a simultaneous decrease of the $S_1$ absorbance, a shift and changes of the $S_2$ peak and an increase of the positive or negative trion peak at 1.061\,eV.
\begin{figure}[htbp]
	\centering
		\includegraphics[width=8.5 cm]{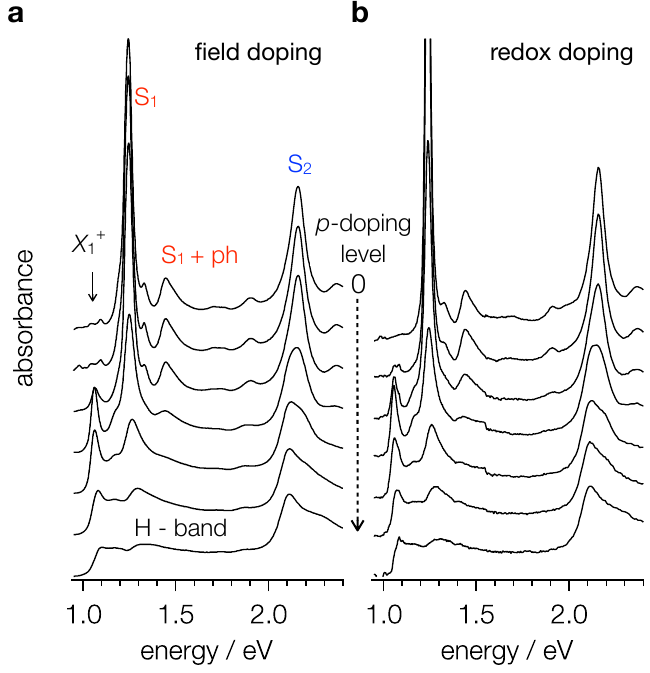}
		\caption{{\bf Comparison of absorption spectra for electrochemically and redox-chemically doped s-SWNTs.} {\bf a)} Selected spectra of a (6,5) enriched film of electrochemically {\it p}-doped s-SWNTs from Fig. \ref{fig3}. {\bf b)} Selected spectra of a (6,5) enriched dispersion of redox-chemically {\it p}-doped s-SWNTs from Fig. \ref{fig2}.}
		\label{fig4}
\end{figure}

A comparison of selected spectra from both datasets reveals a striking similarity, with practically no noticeable difference between redox- and electrochemically field-doped s-SWNTs apart from the slightly broader exciton bands for the thin film sample (see Fig. \ref{fig4}). Such broadening is commonly attributed to intertube coupling in nanotube networks \cite{Crochet2007, Crochet2011}. Even subtle details such as the appearance of a lower energy feature within the second subband exciton range at about 2.12\,eV are nearly indistinguishable for both samples. We take this as evidence for the nature of charge transfer states and the microscopic distribution of charges being similar for both doping schemes. This observation suggests that absorption spectra contain information on doping levels irrespective of the process by which surplus carriers were introduced - a key finding when using absorbance changes as a metrological tool for the assessment of doping.

Along with the emergence of a trion peak, the bleach of the first subband exciton absorption band is one of the primary signatures of increasing doping levels, see Fig. \ref{fig4}. This can easily be explained within the homogeneous as well as the in-homogeneous doping scenarios. For in-homogeneous doping with charge carriers localized in certain nanotube sections, the bleach can simply be attributed to a reduced length of the pristine, i.e. intrinsic nanotube sections that exhibit the unperturbed exciton absorption band. In this case the bleach of the exciton band should be more or less independent of the exciton size. If for example 50\% of the total nanotube length is doped in-homogeneously then the exciton oscillator strength should decrease by a proportionate amount.

\begin{figure}[htbp]
	\centering
		\includegraphics[width=8.5 cm]{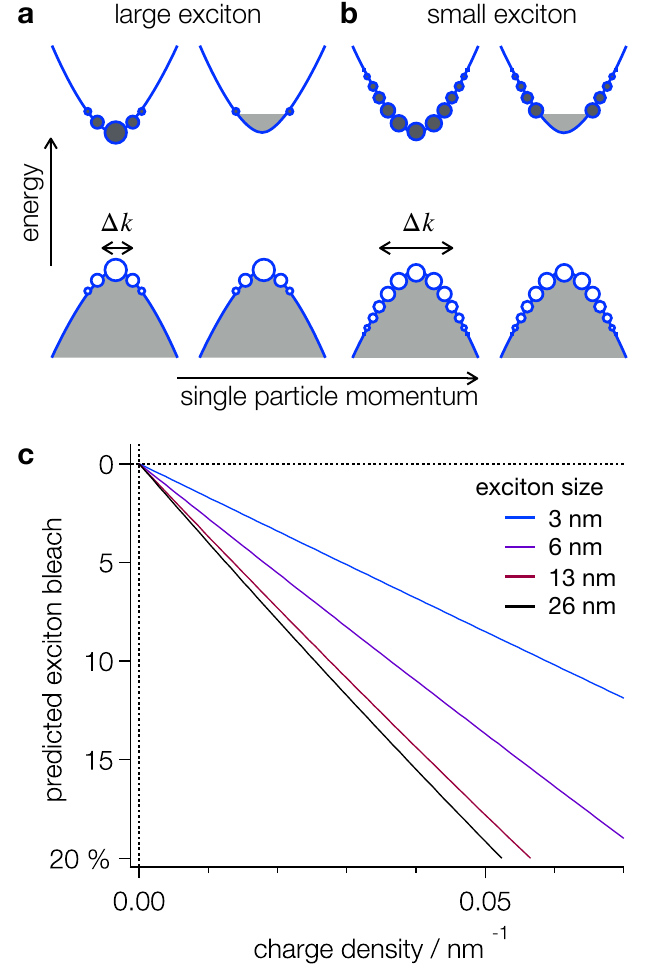}
		\caption{{\bf Exciton bleach and homogeneous doping.} {\bf a)} and {\bf b)} Schematic illustration of the effect of exciton size on the participation of {\it k}-states to exciton wave-functions in intrinsic and doped s-SWNTs. {\bf c)} Calculated decrease of the exciton oscillator strength with surplus carrier concentration as predicted by the phase space filling model for different exciton sizes ($T=293\,{\rm K}$).}
		\label{fig5}
\end{figure}

In contrast, for homogeneous doping the associated change in exciton absorbance is closely associated with the exciton size, as illustrated schematically in Fig. \ref{fig5}a and b. Here, the exciton size is determined by the number of k-states that participate in the formation of the exciton, with a broad distribution of momentum states leading to the formation of small excitons and with the superposition of few momentum states leading to the formation of small excitons - in line with expectations from Heisenberg's uncertainty principle. In other words: larger excitons borrow their oscillator strength from small energy windows at the edges of valence and conduction bands and they are consequently more significantly affected by small doping levels since excess carriers begin by blocking precisely these states (see Fig. \ref{fig5}a and b).      

The magnitude of this effect can be predicted theoretically using a modified phase space filling (PSF) model, based on the work of Schmitt-Rink and coworkers for the interpretation of the nonlinear exciton response to optical excitation \cite{Schmittrink1985}. The PSF model was originally conceived for the determination of exciton Bohr radii from nonlinear optical spectroscopy \cite{Schmittrink1985,Luer2009, Mann2016}. However, in the approach used here, the oscillator strength is not bleached by optical excitation, which simultaneously removes electron and hole states from the coherent superposition of wavefunctions, but instead by surplus charges, leading to a reduction of available states for one carrier type only (for details see supporting information). In line with the treatment by Schmitt-Rink et al. we also neglect screening effects and assume that the exciton bleach is mainly driven by the reduction of available valence ({\it p}-doping) or conduction band states ({\it n}-doping).

The results of these calculations are shown in Fig. \ref{fig5}c. The colored straight lines represent the predicted change of oscillator strength at 293 K, for an effective hole mass of $m_{\rm eff}=0.07\,m_e$ ($m_e$ - electron mass) \cite{Hartleb2015} and an exciton size ranging from $3$ to $26\,{\rm nm}$. Previous experimental findings indicate that the exciton size to be used for (6,5) s-SWNTs is $13\,{\rm nm}$ \cite{Hartleb2015, Mann2016}. These results suggest that carrier concentrations as little as 0.01 $\rm nm^{-1}$, corresponding roughly to $10^{-4}$ on a per carbon atom basis, should lead to easily detectable exciton absorbance changes of a few percent.   

Next we determine the concentration of excess carriers in electrochemically doped s-SWNTs as a function of the applied potential using a band-filling model which implicitly assumes homogeneous doping. In short, this amounts to calculating the carrier density from the nanotube density of states (DOS) $\tilde{g}(E)$, using the interface potential $\phi_q$ and the band gap of 1.55 eV as reported previously for (6,5) s-SWNTs \cite{Hartleb2015}. 

When doing so, care has to be taken in determining the contributions of the electrostatic potential drop $\phi_q$ between working electrode and SWNT as well as the potential drop $\phi_{\rm DL}$ across the diffuse layer to the experimentally measured working electrode potential $U_{\rm WE}=\phi_q+\phi_{\rm DL}$ \cite{Heller2006,Li2018,hertel2018}. This is necessary because the quantum capacitance for charging of s-SWNT is expected to be of similar magnitude as the diffuse double layer capacitance \cite{hertel2018}. The latter can be calculated from the Gouy-Chapman-Stern (GCS) model which thereby allows removing the contribution of the double layer potential $\phi_{\rm DL}$ to $U_{\rm WE}$ \cite{Bard2001}. 

\begin{figure}[htbp]
	\centering
		\includegraphics[width=8.5 cm]{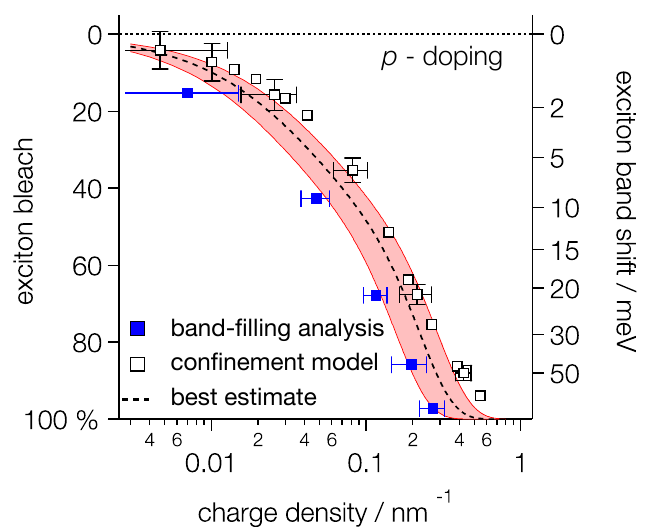}
		\caption{{\bf Dependence of exciton bleach on doping levels.} Surplus carrier concentrations are compared for data analyzed using the band filling model (blue solid squares, field-doped) and for data analyzed using a confinement model (black open squares, redox-doped) by plotting the experimentally measured exciton bleach versus the model-based carrier concentrations. The dashed line represents the best estimate absent a better understanding of carrier localization.}
		\label{fig6}
\end{figure}

The carrier density $n$ in units of elementary charges per unit length is then given by 
\begin{equation}
	n=\int \tilde{g}(E)\,f(E+e\phi_q,T)\,dE
	\label{eq5}
\end{equation} 
where $f(E,T)$ is the Fermi-Dirac distribution at temperature $T$. The DOS is obtained from curvature-corrected tight binding band structure calculations \cite{Hartleb2015, Hagen2004}. The results of this calculation for an electrochemically doped (6,5) s-SWNT thin film sample are shown as blue squares in Fig. \ref{fig6}.  

Alternatively, carrier concentrations can be obtained from a confinement model explicitly assuming an in-homogeneous distribution of charge carriers as discussed by Eckstein {\it et al.} \cite{Eckstein2017}. Charged s-SWNTs in solvent environments attract solvated counterions whose adsorption is associated with the formation of the electrochemical Helmholtz layer. As a point of reference, we calculate the interaction strength between opposing point charges using the relative permittivity $\epsilon=7.43$ of THF, with one charge at the center of a (6,5)- s-SWNT and the other at 0.9\,nm or at 0.7\,nm distance. These values correspond to the approximate spacings between the charge on the nanotube and a counterion with or without solvation shell, respectively \cite{Eckstein2017}. The corresponding Coulomb energies are -95 and -114\,meV, respectively. Note that these interactions not only provide a thermodynamic incentive for the adsorption of counterions on the surfaces of charged SWNTs and thereby for the formation of the Helmholtz layer. These interactions also favor the localization of surplus charges on the nanotube in the vicinity of the adsorbed counterion \cite{Kim2011,Crochet2012}.

Thus it has recently been argued that for {\it p}-doped s-SWNTs such counterion adsorption leads to the formation of split-off valence band states which provide deep charge traps leading to the localization of most surplus charges on the s-SWNT \cite{Eckstein2017}. These charged traps also act as barriers for exciton motion thereby effectively confining exciton wavefunctions to non-doped intrinsic sections of the nanotube. The resulting confinement-induced energy shift can then be used for a determination of the mean spacing between trapped charges, thereby allowing for a quantitative assessment of carrier concentrations \cite{Eckstein2017}.

In Fig.\,\ref{fig6} we compare the correlation between carrier concentrations obtained for electro- and for redox-chemically doped nanotube samples with the associated bleach of the first subband exciton transition. As discussed above, the carrier concentrations are obtained using the homogenous band-filling model (blue squares) and the in-homogeneous confinement model (open squares) \cite{Eckstein2017}. The right axis indicates the associated blue-shift of the exciton band maximum for the redox-doped sample.

Both models suggest a qualitatively similar dependence of the exciton bleach on the calculated charge density from about 0.005 to 0.5 $\rm nm^{-1}$ with the confinement model systematically yielding somewhat higher values. The red shaded confidence band in Fig.\,\ref{fig6} represent an uncertainty of $\pm 30\%$ with respect to the best estimate obtained from both datasets (dashed black line). Note, that a confinement model analysis of the electrochemically doped s-SWNT sample would yield higher carrier densities, essentially in line with those obtained for the redox-doped s-SWNT sample. 

The precision with which carrier concentrations can in principle be determined from optical spectra should thus be significantly higher than suggested by the spread of the results using the band-filling and confinement models. This is because the limited accuracy of the present analysis is largely due to the uncertainty regarding the degree of carrier localization or delocalization, an issue which should be addressed by future experiments.       

\section{Summary and Conclusions}

We determined carrier concentrations from NIR-VIS absorption spectra of electro- and redox-chemically {\it p}-doped s-SWNT samples using a band filling- and confinement model, respectively. Spectral changes appear to be independent of the specific scheme used for doping. The key spectral signature used for assessing carrier densities is the change of the first subband exciton oscillator strength. Its sensitivity to doping levels within a homogeneous doping scenario is predicted using a phase-space filling (PSF) model. The phase space filling model predicts that large excitons are more sensitive to doping than smaller excitons. At carrier concentrations up to about 0.1 $\rm nm^{-1}$ and an exciton size of 13 nm \cite{Mann2016} this model yields $n=-0.28\,{\rm nm^{-1}}\times \Delta f/f$.

The confinement model analysis of experimental data from redox doped (6,5) s-SWNTs on the other hand assumes that surplus carriers are in-homogeneously distributed. In this case the exciton bleach is characterized by a nearly linear dependence on surplus carrier concentrations up to about 0.2 $\rm nm^{-1}$ with $n=-(0.27\pm0.02)\,{\rm nm^{-1}}\times \Delta f/f$. The seemingly good agreement between the predictions of the PSF model and the experimental analysis using the confinement model is most likely somewhat coincidental.

Nonetheless further improvement of the precision with which carrier concentrations can be obtained from changes in exciton oscillator strengths should be possible, given a better understanding of the character of the distribution of surplus charges in s-SWNTs. Specifically the issue of carrier localization in deep charge traps appears to warrant further investigation. 

\begin{suppinfo}
Description of the phase space filling model.
\end{suppinfo}

\section{Acknowledgements}
T.H. and K.E. acknowledge financial support by the German National Science Foundation through the DFG GRK2112. T.H. and M.A. also acknowledge financial support through DFG grant HE 3355/4-1. T.H. acknowledges support through the Alumni program of the Alexander von Humboldt Foundation.

\end{document}